\documentclass[12pt, preprint]{aastex}

\usepackage{subfigure}
\usepackage{color}
\usepackage{hyperref}
\usepackage{url}
\usepackage{natbib}
\usepackage{amsmath}

\newcommand{\project}[1]{\textsl{#1}}
\newcommand{\cpm}{\project{CPM}}
\newcommand{\cpmdiff}{\project{CPM Difference Imaging}}

\newcommand{\kepler}{\project{Kepler}}

\newcommand{\KTCN}{\project{K2 Campaign 9}}
\newcommand{\epic}{\project{EPIC}}

\newcommand{\set}[1]{\mathcal{#1}}

\graphicspath{{figures/}}

\definecolor{linkcolor}{rgb}{0,0,0.5}
\hypersetup{colorlinks=true,linkcolor=linkcolor,citecolor=linkcolor,
            filecolor=linkcolor,urlcolor=linkcolor}

\begin{document}

\title{A pixel-level model for event discovery in time-domain imaging}
\author{%
  Dun~Wang\altaffilmark{\ref{CCPP},\ref{email}},
  David~W.~Hogg\altaffilmark{\ref{CCPP},\ref{CDS},\ref{MPIA},\ref{FI}},
  Daniel~Foreman-Mackey\altaffilmark{\ref{UW},\ref{SF}}
  Bernhard~Sch\"olkopf\altaffilmark{\ref{MPIIS}}
  }
\newcounter{address}
\setcounter{address}{1}
\altaffiltext{\theaddress}{\stepcounter{address}\label{CCPP}%
  Center for Cosmology and Particle Physics, Department of Physics, New York University}
\altaffiltext{\theaddress}{\stepcounter{address}\label{email}%
  To whom correspondence should be addressed; \texttt{<dun.wang@nyu.edu>}.}
\altaffiltext{\theaddress}{\stepcounter{address}\label{CDS}%
  Center for Data Science, New York University}
\altaffiltext{\theaddress}{\stepcounter{address}\label{MPIA}%
  Max-Planck-Institut f\"ur Astronomie, Heidelberg, Germany}
\altaffiltext{\theaddress}{\stepcounter{address}\label{FI}%
  Flatiron Institute, Simons Foundation}
\altaffiltext{\theaddress}{\stepcounter{address}\label{UW}%
 Astronomy Department, University of Washington, Seattle, WA 98195}
\altaffiltext{\theaddress}{\stepcounter{address}\label{SF}%
NASA Sagan Fellow}
\altaffiltext{\theaddress}{\stepcounter{address}\label{MPIIS}%
  Max-Planck-Institut f\"ur Intelligente Systeme, T\"ubingen}

\begin{abstract}
Difference imaging or image subtraction is a method that measures differential photometry by matching the pointing and point-spread function (PSF) between image frames. 
It is used for the detection of time-variable phenomena.
Here we present a new category of method---\cpmdiff, in which differences are not measured between matched images but instead between image frames and a data-driven predictive model that has been designed only to predict the pointing, PSF, and detector effects but not astronomical variability. 
In \cpmdiff\ each pixel is modelled by the Causal Pixel Model (\cpm) originally built for modeling \kepler\ data, in which pixel values are predicted by a linear combination of other pixels at the same epoch but far enough away such that these pixels are causally disconnected, astrophysically. 
It does not require that the user have any explicit model or description of the pointing or point-spread function of any of the images.
Its principal drawback is that---in its current form---it requires an imaging campaign with many epochs and fairly stable telescope pointing.
The method is applied to simulated data and also the \project{K2 Campaign 9} microlensing data. 
We show that \cpmdiff\ can detect variable objects and produce precise differentiate photometry in a crowded field.
\cpmdiff\ is capable of producing image differences at nearly photon-noise precision. 
\end{abstract}


\section{Introduction}

\subsection{Difference imaging}
Difference imaging or image subtraction is a method developed for detecting variable objects in astronomical studies, in which the difference is measured between two images that are both positionally and photometrically matched. This kind of method is optimal for analyzing variability in crowded astronomical images, since it bypasses the procedure of doing photometry for each individual object and comparing with a catalog, but instead directly measures differential photometry.
The common workflow for difference imaging is:
\begin{enumerate}
\item
Create a reference image by either stacking image frames or selecting the frame with the best seeing.
\item
Astrometrically register each frame to the reference frame.
\item
Match the seeing between each image and the reference frame by fitting a convolution kernel that accounts for the differences between the point spread functions(PSFs).
\item
Match the mean throughput or photometric calibration of the two frames and subtract to get a difference image.
\end{enumerate}
The main challenge of the difference imaging problem lies in the inference of the convolution kernel that corrects the difference of PSFs between two frames.
The first attempt at difference imaging or image subtraction was made by \cite{imagesub1}, who calculated a convolution kernel by taking the ratio of two images of a bright star in Fourier space. 
This method is straight-forward, but it is numerically unstable and sensitive to noise.
\cite{alard} improved the method by decomposing the kernel into a linear combination of basis functions and then fitting a constant convolution kernel to match the PSFs of images.
The current preference for difference imaging \citep{varyingkernel} is to divide images into sub-areas and fit a varying kernel to account for the spatial variation of the PSF. 
This method is implemented and widely used as \project{HOTPANTS}\footnote{\url{http://www.astro.washington.edu/users/becker/v2.0/hotpants.html}} and \project{ISIS}\footnote{\url{http://www2.iap.fr/users/alard/package.html}}, \cite{varyingkernel}. 
Although \cite{alard}, \cite{varyingkernel} mitigate the numerical instability, the choice of basis function significantly affects the performance and may require sufficient information about the PSFs in both the reference and target images.
\cite{bramich} handles more complicated kernels by using a discrete pixel array instead of a linear combination of basis functions.
The pixelized kernel increases the flexibility. 
However it is much easier to overfit and more sensitive to noise. 
As pointed in \cite{regularization}, regularization on smoothness and compactness of the kernel is essential.  
\cite{optimal} based on the likelihood ratio test and derived a closed form for the subtraction image, in which the target and reference images are convolved with each other's PSF.
Instead of solving the convolution kernel, good PSF estimation for both the reference and new images is required.

Most of these methods also require very precise image registration. 
In \cite{alard} and \cite{varyingkernel}, due to the restriction of the basis functions, the kernel is not able to model complex transformations between the reference and target. 
In \cite{optimal}, the convolution procedure can only match the difference of PSFs, making precise astrometry is crucial.
The requirement for good registration and PSF estimation makes the method perform worse with under-sampled data.
Finally, all these methods require to construct a reference image. 
The choice of the reference image is important for both kernel solving and subtraction, but no robust algorithm for reference image selection has been proposed so far \citep{reference}.

The difference imaging methods discussed so far \citep{imagesub1, alard, varyingkernel, bramich} all follow the same framework; the only difference lies in how the convolution kernel is calculated, so hereafter in this paper we refer to these methods as classical difference imaging. 
\cite{optimal} does not explicitly solve for a convolution kernel, but they still follow the same basic procedure, so we categorize the method into the classical approach.
The method proposed in this paper, \cpmdiff, is different from the classical approach, because it does not require a reference image. 
Instead of modelling each image using a reference, each pixel is modelled directly as a linear combination of other pixels from the same image. 
\cpmdiff\ does not explicitly model the PSF of any image
or difference of images. 
Instead, it relies on the assumption that changes in the pointing and PSF will affect all pixels in the same field of view, although not necessarily in the same way. 
In \cpmdiff, images are not compared to a reference image in isolation. 
Instead, an optimized estimate of the reference value for every pixel is computed using different pixels and measurements at different times.

The benefits from the proposed method in this paper can be listed as: 
\begin{enumerate}
\item No knowledge of the detailed pointing or PSF of any frame is required. 
\item The method performs well on under-sampled imaging
\end{enumerate}

Classical difference imaging has been successfully applied for the detection of variable sources in microlensing \citep{macho, ogle} and supernova \citep{sdss} surveys.
These methods will play an important role in time-domain astronomy in the near future. 
\project{LSST} \citep{lsst} will image the entire night sky repeatedly to find distant transient events of known kinds and even discover new classes of variable objects, \project{TESS} \citep{tess} will produce a continuous series of full frame images covering 2300 $deg^2$ of the sky with 30-minute cadence, in which a lot of variable sources such as exoplanets, near-Earth asteroids, bright AGN outbursts and nearby supernovae will be detected and \project{Euclid} \citep{Euclid} and \project{WFIRST} \citep{wfirst} are likely to have time-domain survey fields.

\section{The Method} \label{method}
\cpmdiff\ requires multiple images of the same field with registration better than about one PSF width.
This ensures that the corresponding pixels from different images are generally illuminated by the same sources. 
More precisely, the assumption is that the variations in the measured scene due to shifts, rotations, and PSF variations are small enough to be treated in with reasonable accuracy in a quasi-linear regime.

In \cpmdiff, each pixel is modeled, and the difference is measured between the model and the data. 
The details of the model are almost identical to the \cpm\ model \citep{cpm} and more detail about what assumptions are being made are also spelled out there. 
The theoretical analysis of the \cpm\ model can be found in \cite{pnas}.
Here we briefly outline the basic procedure. 
Each pixel value $I_{m,n}$ of pixel $m$ at time $t_n$ is predicted by a linear combination of pixel values $I_{m',n}$, where $m'$ is from a set of pixels $m'\in\set{M}_m$ that are on the same CCD but far enough away from the target pixel $m$ to not be significantly illuminated by the same source. 
This model can be written as,
\begin{eqnarray}
I_{m,n}         &=& I^{\ast}_{m,n} + e_{m,n}
\\
I^{\ast}_{m,n}  &=& \sum_{m'\in\set{M}_m} a_{m,m'}I_{m',n} 
\quad,
\end{eqnarray}
where $I^{\ast}_{m,n}$ is the model prediction (by the model) for data point $I_{m,n}$, $e_{m,n}$ is the residual away from the prediction, and the $a_{m,m'}$ are parameters (linear coefficients of the prediction).
Assuming Gaussian noise, the parameters $a_{m,m'}$ are estimated by standard $\chi^2$ minimization with an additional regularization term that penalizes large squared values for the coefficients $a_{m,m'}$:
\begin{eqnarray}
\chi^2_{m}    &=& \sum_{n} \frac{[I_{m,n} - I^{\ast}_{m,n}]^2}{\sigma^2_{m,n}}+ \lambda_{a}\sum_{m'\in\set{M}_m}a_{m,m'}^2 
\quad,
\end{eqnarray}
where the $\sigma^2_{m,n'}$ are the individual-pixel noise variances, and $\lambda_{a}$ set the strength of the regularization for parameters $a_{m,m'}$.
Since the model for $I^{\ast}$ is linear, the optimized coefficients can be computed analytically.

In \cpm, the number of predictor pixels in $\set{M}_m$ and the regularization strength $\lambda_{a}$ are two hyper-parameters that need to be set by cross-validation to optimize the performance. 
In addition, the method we employ to choose the predictor pixels should also be explored and optimized.
Since finding the optimal $\set{M}_m$ is complicated and time-consuming, in this paper we excluded 10 closest rows and columns from the target pixel and from the remaining pixels, we chose the 400 closest pixels that are at least 16 pixels away from the target pixel.
With the $\set{M}_m$ settled, $\lambda_{a}$ was chosen by running a coarse-grid cross-validation. 
Here the setting of the hyper-parameters is not optimal and only for demonstration. 
We will talk more about the hyper-parameters and ranking of predictor pixels in the discussion section.

With the modelled pixel values $I^{\ast}_{m,n}$, the difference image is defined as the difference between the model and the data:
\begin{eqnarray}
D_{m,n} &=& I_{m,n} - I^{\ast}_{m,n}
\quad.
\end{eqnarray}

\section{Experiments}
In order to illustrate the performance \cpmdiff, we present several experiments on mock and real data. 
First, \cpmdiff\ is tested under different observation conditions (space-based and ground-based) with mock data. 
Then,  the method is applied to the \KTCN\ data to show how it performs with real data. 
In the end of this section,  large variations of pointing and PSF are tested with mock data to study the limitations of the method.

\subsection{Mock data}
To produce the mock image, \project{TRILEGAL\footnote{\url{http://stev.oapd.inaf.it/cgi-bin/trilegal}}} \citep{TRILEGAL}  is used to generate a catalog of stars with magnitudes. 
The initial coordinates of the stars are randomly drawn from a 2-d uniform distribution. 
For each frame of the image, the same affine transformation is applied  to all the stars to imitate the pointing motion and rotation of the camera:
\begin{eqnarray}\label{transformation}
\begin{bmatrix}
    x' \\
    y' \\
    1
\end{bmatrix}
&=&
\begin{bmatrix}
    \cos \theta & \sin \theta & t_x \\
    -\sin \theta & \cos \theta & t_y \\
    0 & 0 & 1 \\
\end{bmatrix}
\begin{bmatrix}
    x \\
    y \\
    1
\end{bmatrix}
\end{eqnarray}
where $t_x, t_y$ are the translations in x and y direction and $\theta$ is the angle of the rotation.\\
For pixel centered on (r, s), the value of the pixel is evaluated by the equation:
\begin{eqnarray}
p_{r,s} &=& \sum_{i}^{N} PRF(r-x_i, s-y_i) f_i
\end{eqnarray}
where $(x_i,y_i)$ is the coordinate of the star i on the image and $f_i$ is the flux associated with that star. 
$PRF(\Delta x, \Delta y)$ is the pixel response function (or pixel-convolved point spread function). 
Here a 2-d gaussian is used:
\begin{eqnarray} \label{prf}
PRF(\vec{r}) &=& A \exp(-\frac{1}{2} \vec{r}\cdot V^{-1}\cdot \vec{r}) \\
V^{-1} &=& 
\begin{bmatrix}
    \frac {\cos ^{2}\phi }{f_{x}^{2}}+\frac {\sin ^{2}\phi }{f_{y}^{2}} & -\frac {\sin 2\phi }{2f_{x}^{2}}+\frac {\sin 2\phi }{2f_{y}^{2}}  \\
    -\frac {\sin 2\phi }{2f_{x}^{2}}+\frac {\sin 2\phi }{2f_{y}^{2}} & \frac {\sin ^{2}\phi }{f_{x}^{2}}+\frac {\cos ^{2}\phi }{f_{y}^{2}} \\
\end{bmatrix}
\end{eqnarray}
where $\vec{r}$ is the column vector and V is a tensor describing the variance of the PRF.
$f_x$ and $f_y$ are the full width half maximum of the gaussian in x and y direction, and $\phi$ determines the orientation of the gaussian.
Parameters $f_x$, $f_y$, $\phi$ are adjusted to change the shape and width of the PRF.

After the series of images is generated, a flat-field error $\epsilon_{r,s}$ is drawn from a normal distribution with $\mu=1, \sigma=0.01$ to account for the inter-pixel variation of the detector.
With the flat-field error included, the value of the each pixel is $p'_{r,s} = p_{r,s}\epsilon_{r,s}$.
Finally photon noise approximated by a Gaussian with $\sigma = 10^{-4}$ is also added to each individual pixel.

\subsection{Mock space-based data}
\begin{figure}[p]
\begin{center}
\includegraphics[width=0.95\textwidth]{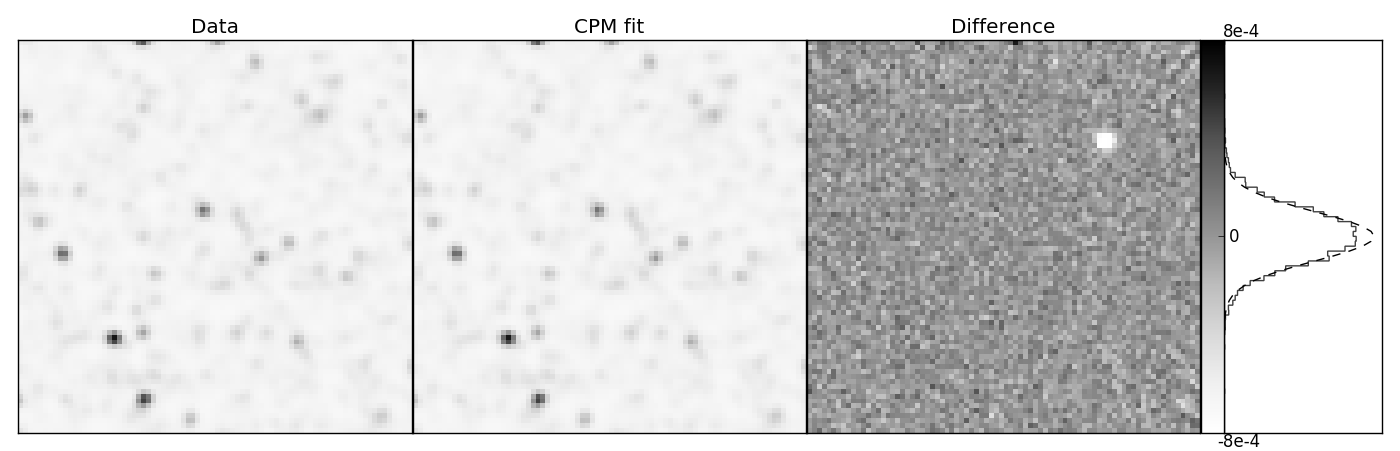}
\includegraphics[width=0.95\textwidth]{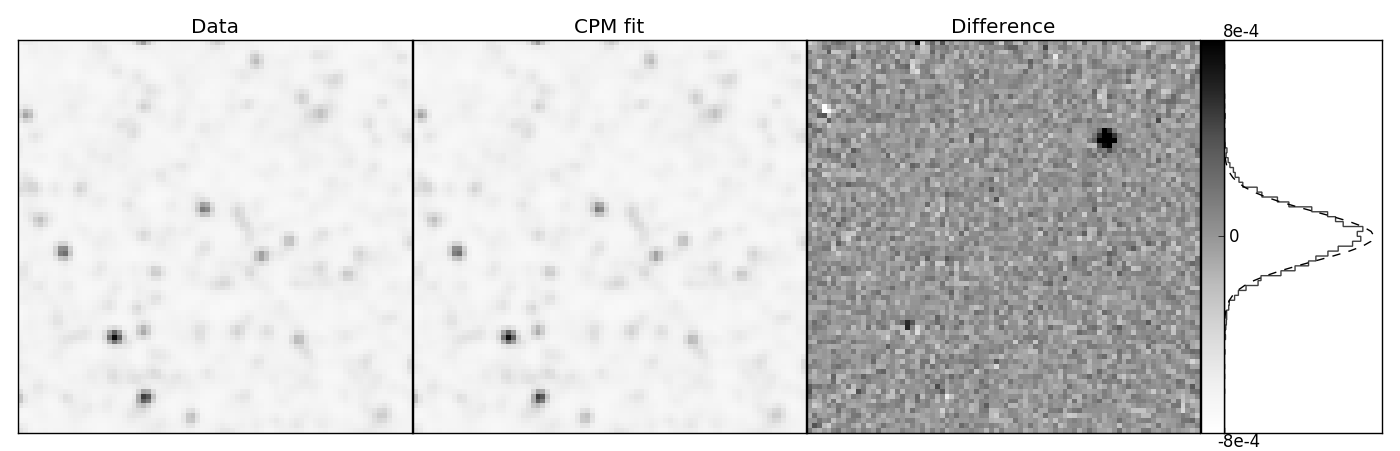}
\includegraphics[width=0.95\textwidth]{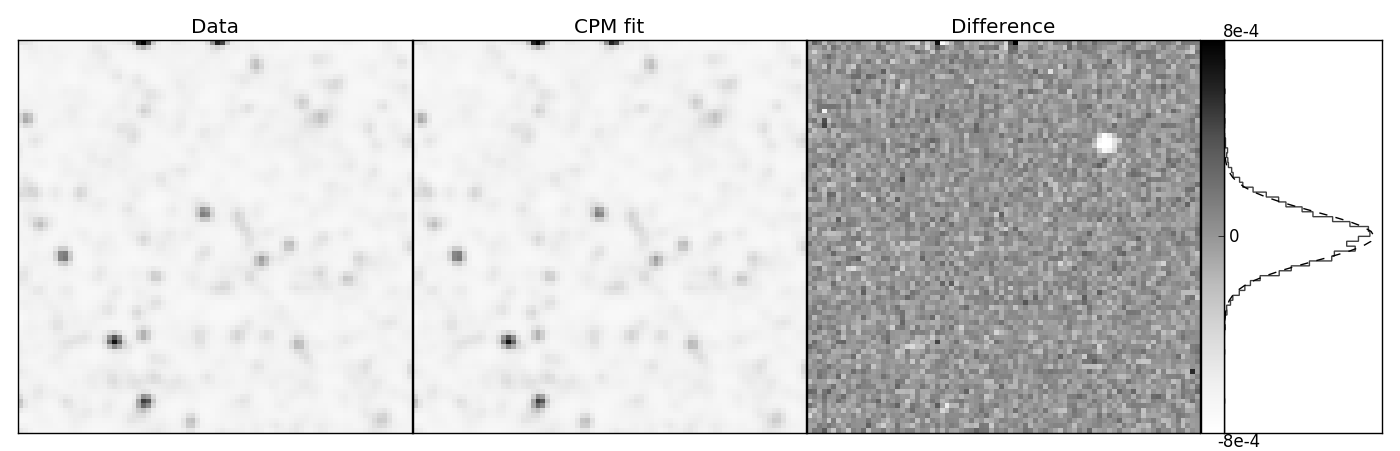}
\end{center}
\caption{
\label{space}
  Mock Space-Based Data---an $80\times 80$ pixel mock data image patch with pointing motion and rotation variation. 
  From top to the bottom,  each row shows a snapshot from different times.
  \emph{Left:} mock data image;
  \emph{Middle:} the prediction of the \cpmdiff;
  \emph{Right:} the relative difference between the data and the prediction, the color bar shows the relative difference;
  the histogram shows the distribution of the difference and the dashed curve is the photon noise: Gaussian with $\sigma = 10^{-4}$.
  Note the small but significant astrometric shifts between images.
  Most of the difference-image pixel values are near zero, except for the variable source (upper-left corner), which shows that \cpmdiff\ can predict the image data, while detecting the variables. 
}
\end{figure}

\begin{figure}[p]
\begin{center}
\includegraphics[width=0.95\textwidth]{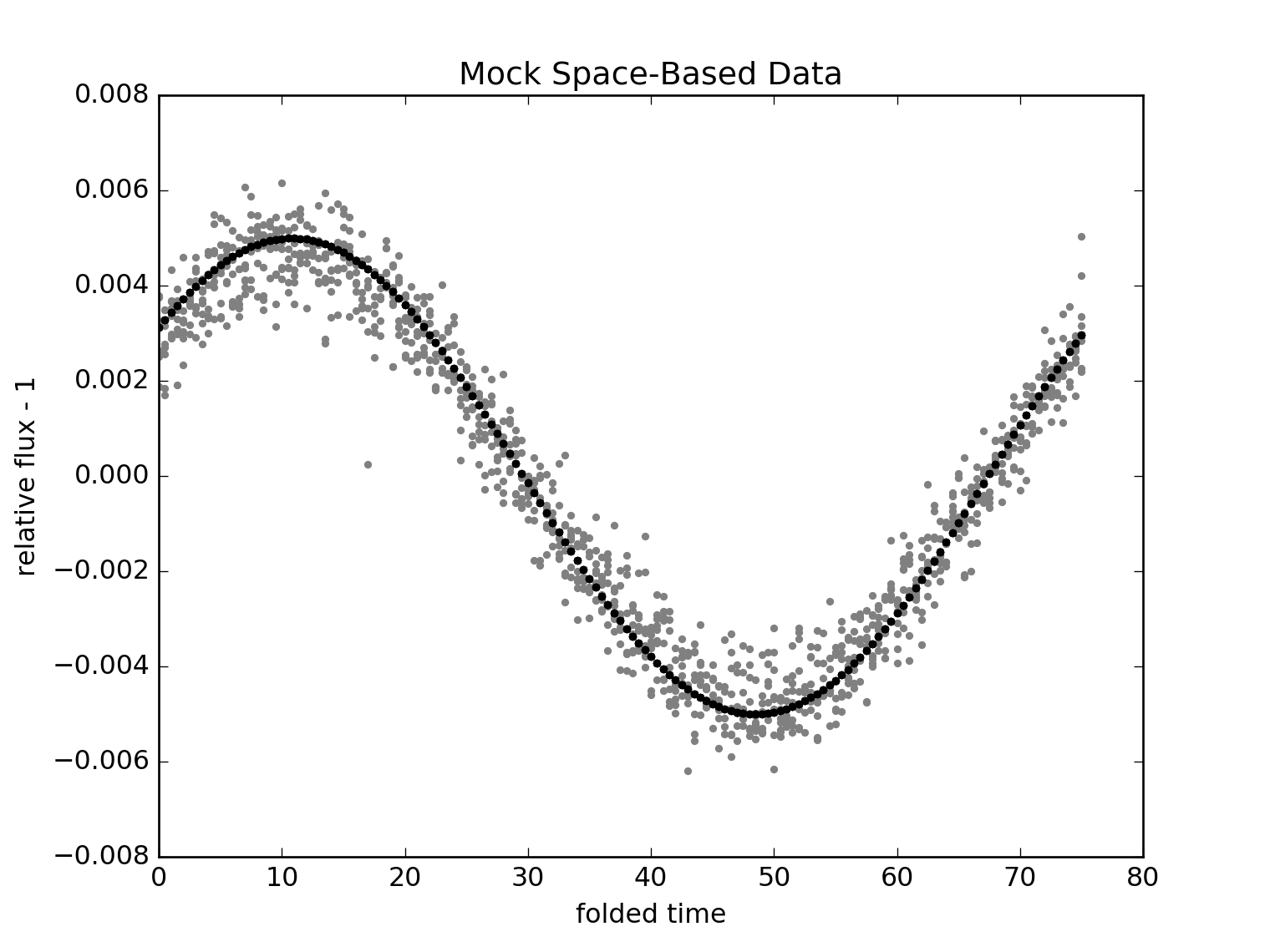}
\end{center}
\caption{
\label{space_lc}
 The folded light curve of the variable star from the mock space-based data shown in Fig.~\ref{space}.
 The black points are the injected signal and the grey points are the recovered light curve from the \cpmdiff\ by co-adding the pixels in a $7\times 7$ patch around the source star.
}
\end{figure}

In space-based observations like those made in the \kepler\ mission, the systematics are mainly caused by changes of the pointing and rotation of the camera, while the PSF is relatively stable. 
Thus the mock space-based data only includes the pointing motion and rotation, but maintains the PSF unchanged from image to image.
In this experiment,  both $t_x $ and $t_y$ in equation \ref{transformation} are drawn from uniform distribution ${\mathcal {U}}(0,1)$ pixels and $\theta$ is drawn from ${\mathcal {U}}(0,0.5)$ deg. 
To illustrate how variable sources will be detected in \cpmdiff, one single periodic variable star (modelled by a constant plus a sine function) is injected in the image.
Fig.~\ref{space} shows three snapshots of the mock data and the \cpmdiff\ of different times. 
The \cpm\ modelled the data with precision close to the photon noise ($10^{-4}$), except the injected variable was still preserved in the difference image. 
Fig.~\ref{space_lc} shows the light curve of the injected variable star and the recovered light curve by co-adding the pixels in a $7\times 7$ patch around the source star.
The choice of the aperture size is not optimal but big enough to include all the flux from the source star. 
There is no PSF and flat-field information used in extracting the light curve. 
Therefore the photometry is not optimal and is only intended for demonstration.

\subsection{Mock ground-based data}
\begin{figure}[p]
\begin{center}
\includegraphics[width=0.95\textwidth]{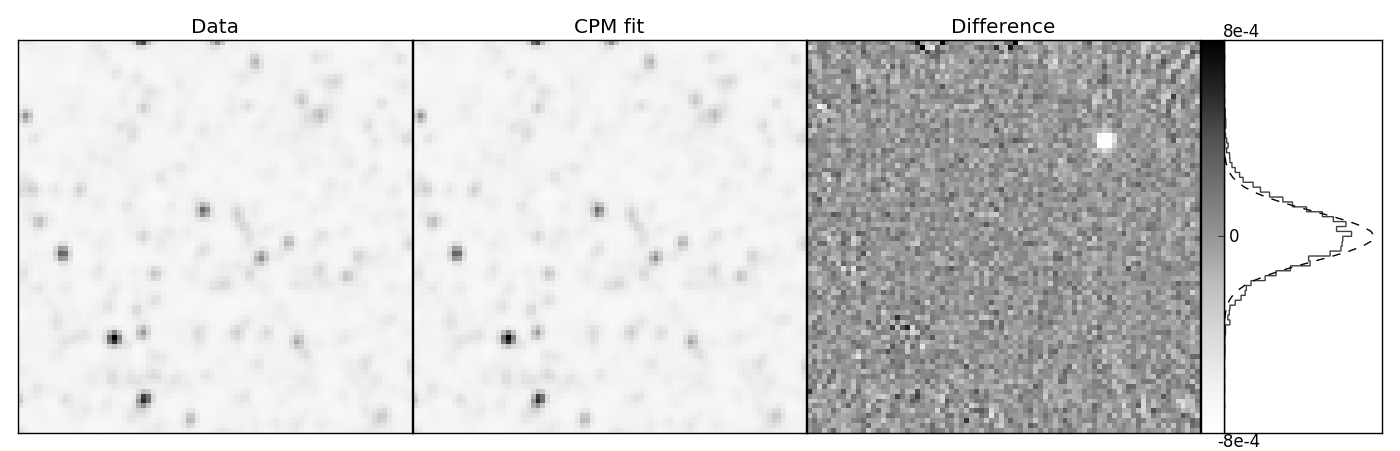}
\includegraphics[width=0.95\textwidth]{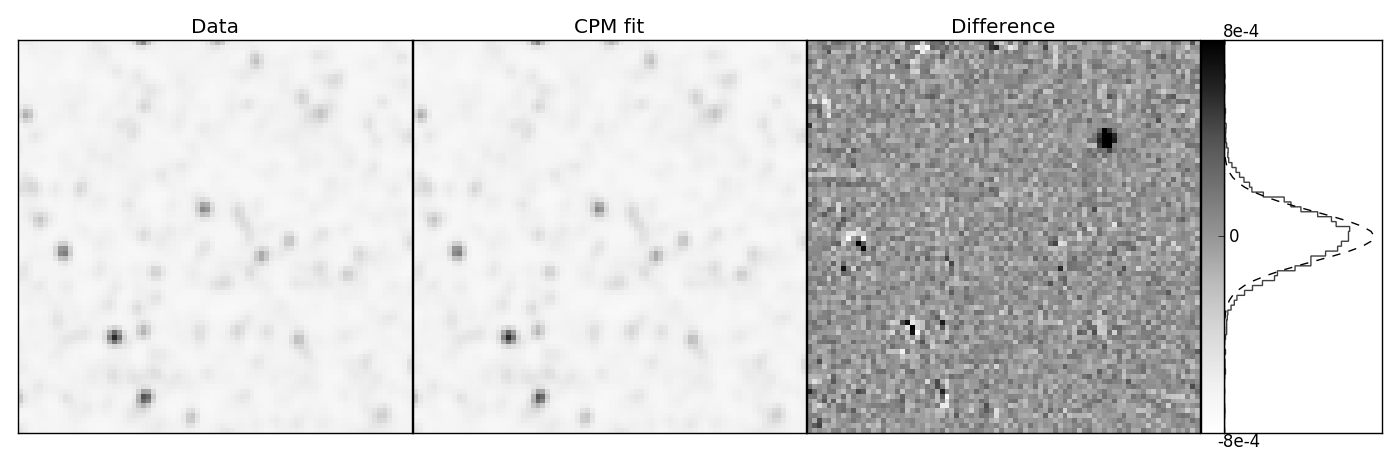}
\includegraphics[width=0.95\textwidth]{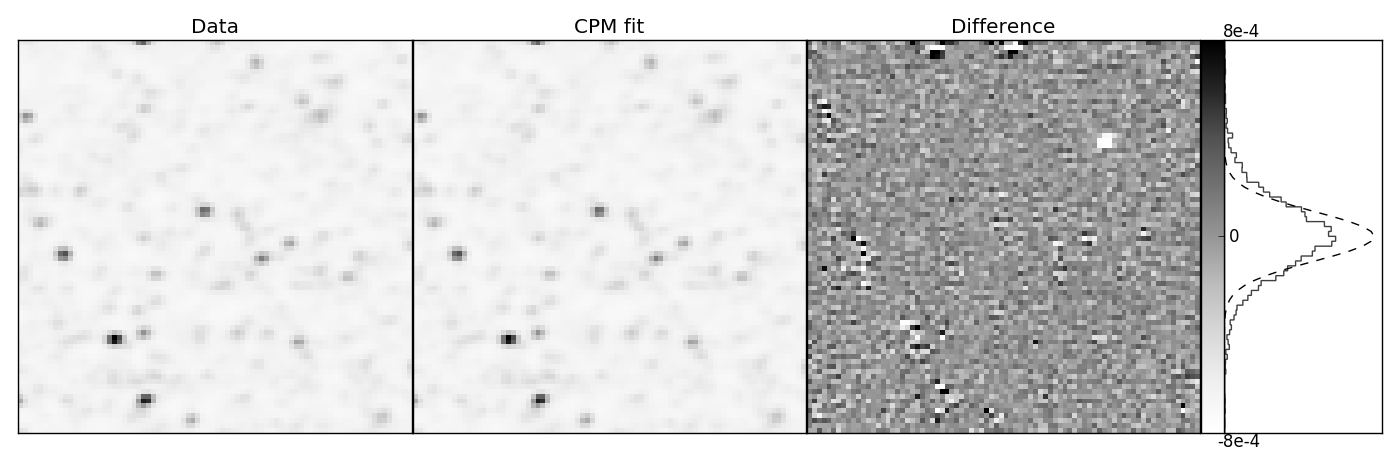}
\end{center}
\caption{
  \label{ground}
  Mock Ground-Based Data---an $80\times 80$ pixel mock data image patch with pointing motion, rotation and PRF variation. 
  From top to the bottom,  each row shows a snapshot from different times.
  \emph{Left:} mock data image;
  \emph{Middle:} the prediction of the \cpmdiff;
  \emph{Right:} the relative difference between the data and the prediction, the color bar shows the relative difference; 
  the histogram shows the distribution of the difference and the dashed curve is the photon noise: Gaussian with $\sigma = 10^{-4}$. 
  As in the space-based test, \cpmdiff\ subtratcted all the constant stars and retained the variable sources with the mock ground-based data, which shows that the method can handle pointing motion, rotation and PRF variation altogether. 
}
\end{figure}

\begin{figure}[p]
\begin{center}
\includegraphics[width=0.95\textwidth]{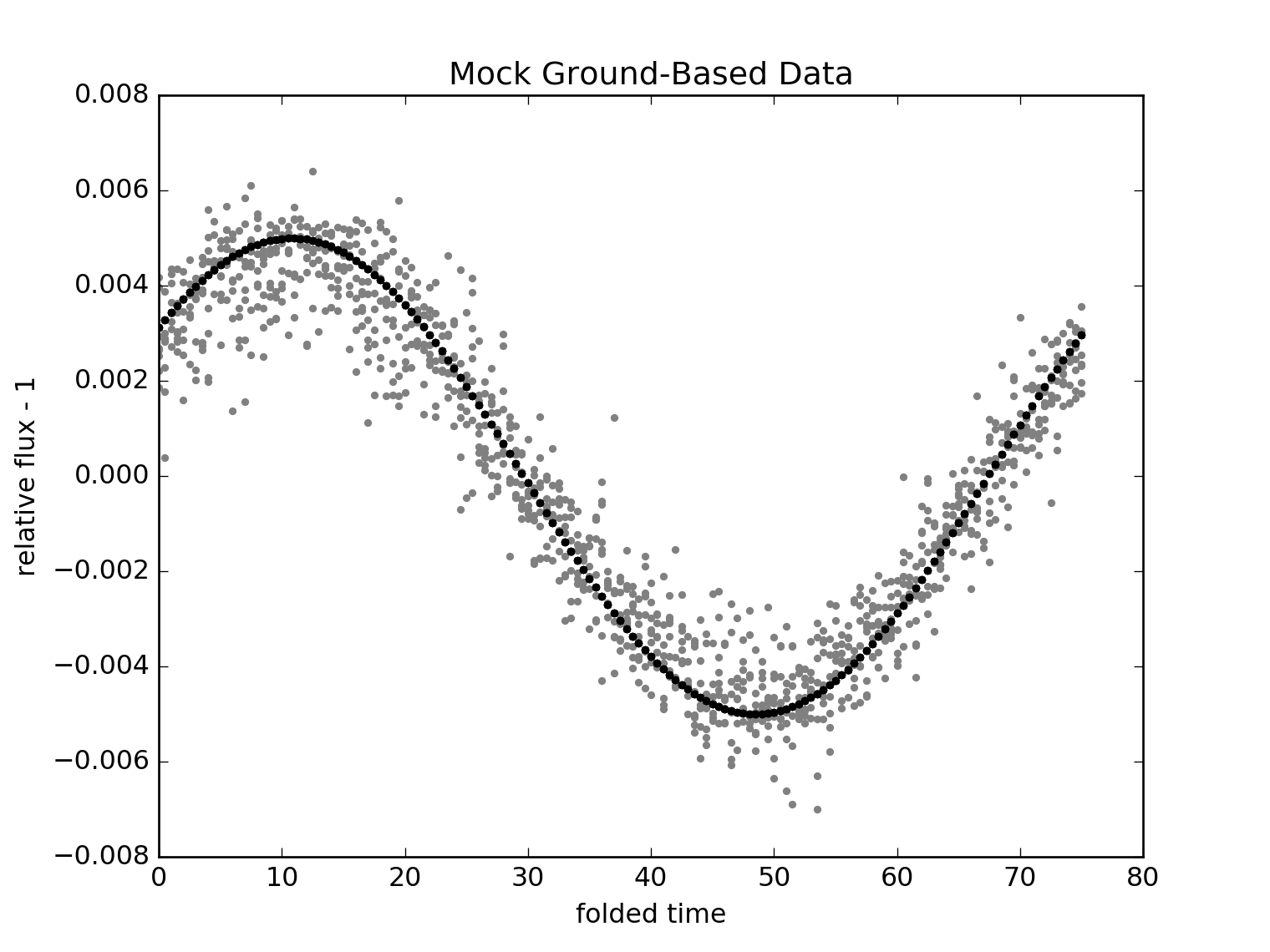}
\end{center}
\caption{
\label{ground_lc}
 The folded light curve of the variable star from the mock ground-based data.
 The black points are the injected signal and the grey points are the recovered light curve from the \cpmdiff\ by co-adding the pixels in a $7\times 7$ patch around the source star.
}
\end{figure}

In ground-based observations, in addition to changes in pointing and rotation, weather changes and atmospheric distortions will also affect the PSF. 
Therefore PSF variations were also included in the mock ground-based data. 
The pointing motion and rotation of the mock data are same as in the space-based test.
Variation of the PSF is achieved by varying the parameters $f_x, f_y, \phi$ defined in equation \ref{prf}.
Both $f_x$ and $f_y$ are drawn from uniform distribution ${\mathcal {U}}(2,3)$ pixels to restrict the full width half maximum of the PRF in both direction within 2-3 pixels and $\phi$ is drawn from ${\mathcal {U}}(0,\pi)$, which allows the orientation of the PRF to be in any direction.
Fig.~\ref{ground} and Fig.~\ref{ground_lc} show the same mock data and light curves as in space-based test, but with PRF variation included.  
As in the space-based test, the \cpmdiff\ was able to model both pointing motion and PRF variations, while still detecting the variable star in the differencing image.
Note that the PRF variations do degrade the quality of the difference image a little, since the changes of the PRF will change the correlation between pixels.
This experiment further confirms that \cpmdiff\ can calibrate both space-based and ground-based data.

\subsection{\KTCN\ real data}
Now we test our method on a $64\times50$ pixel patch (\epic\ 200069960) from \KTCN\footnote{\url{https://keplerscience.arc.nasa.gov/k2-c9.html}}\citep{k2c9}, which was dedicated to a study of gravitational microlensing events.
This data set is an ideal test bed for difference imaging, since it observed a very crowded field near the bulge, where high precision photometry is difficult to achieve directly.
\cite{wei} have already applied the classical difference imaging on this data set and are able to model some microlensing events.

\begin{figure}[p]
\begin{center}
\includegraphics[width=0.95\textwidth]{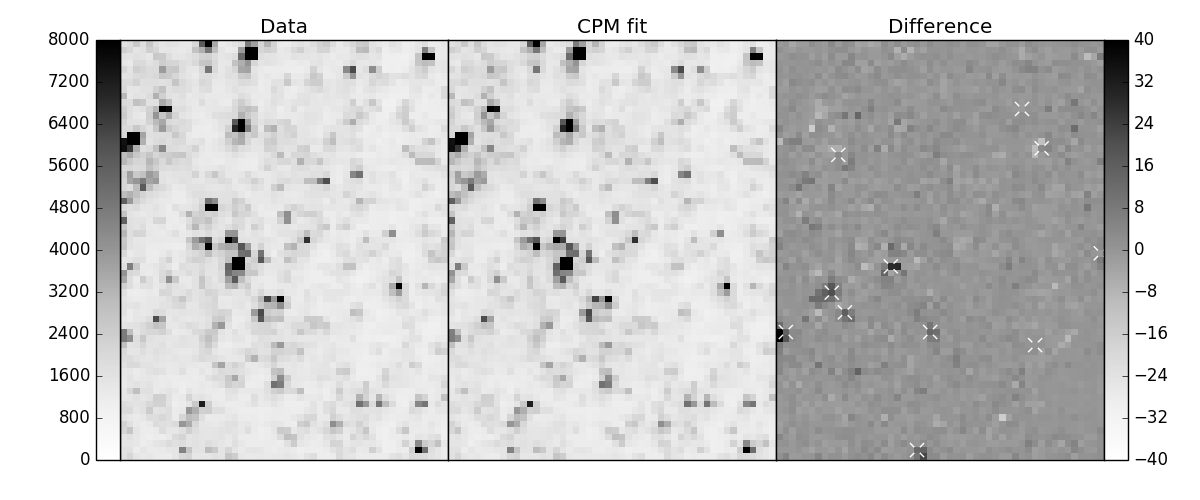}
\includegraphics[width=0.95\textwidth]{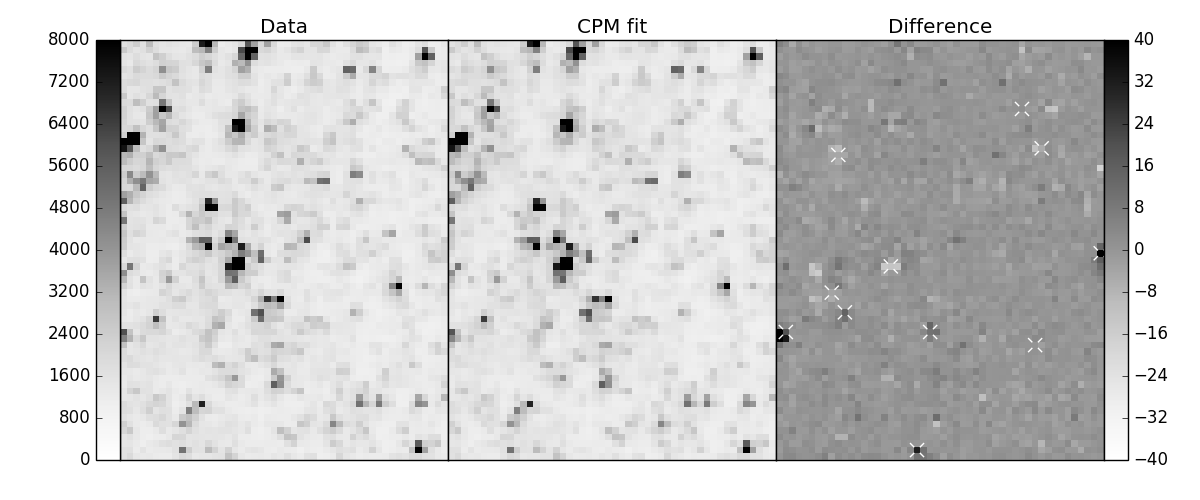}
\includegraphics[width=0.95\textwidth]{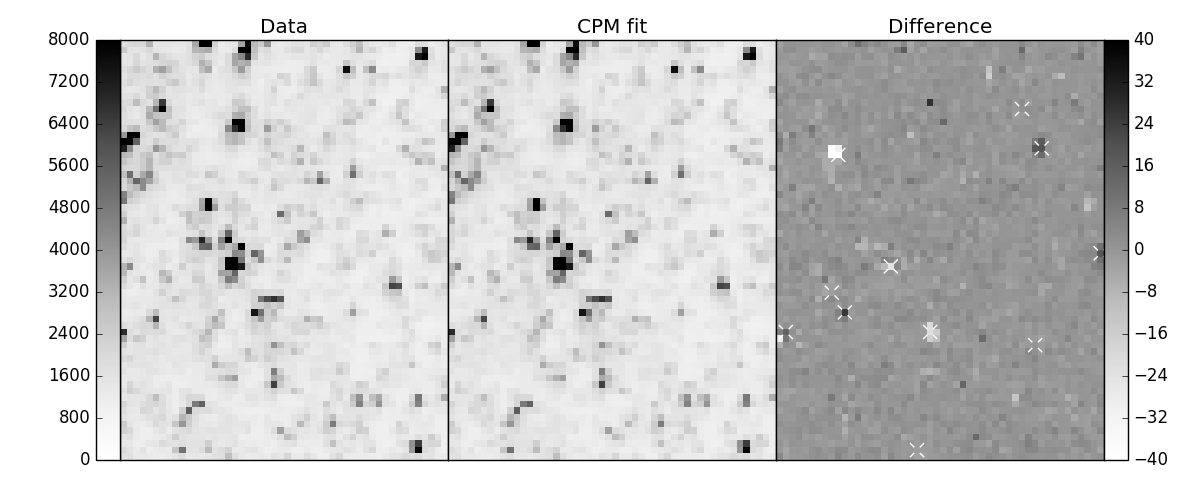}
\end{center}
\caption{
  \label{k2c9}
  \KTCN\ Data---a $64\times 50$ pixel image patch from \KTCN\ (\epic\ 200069960). 
  From top to the bottom,  each row shows a snapshot from different times.
  \emph{Left:} data image;
  \emph{Middle:} the prediction of the \cpmdiff;
  \emph{Right:} the difference between the data and the prediction, white crosshairs indicate detected variable sources.
  The \cpmdiff\ subtracted all the constant sources in the data image, while preserve the variable sources in the difference image.
}
\end{figure}

\begin{figure}[p]
\begin{center}
\includegraphics[width=0.48\textwidth]{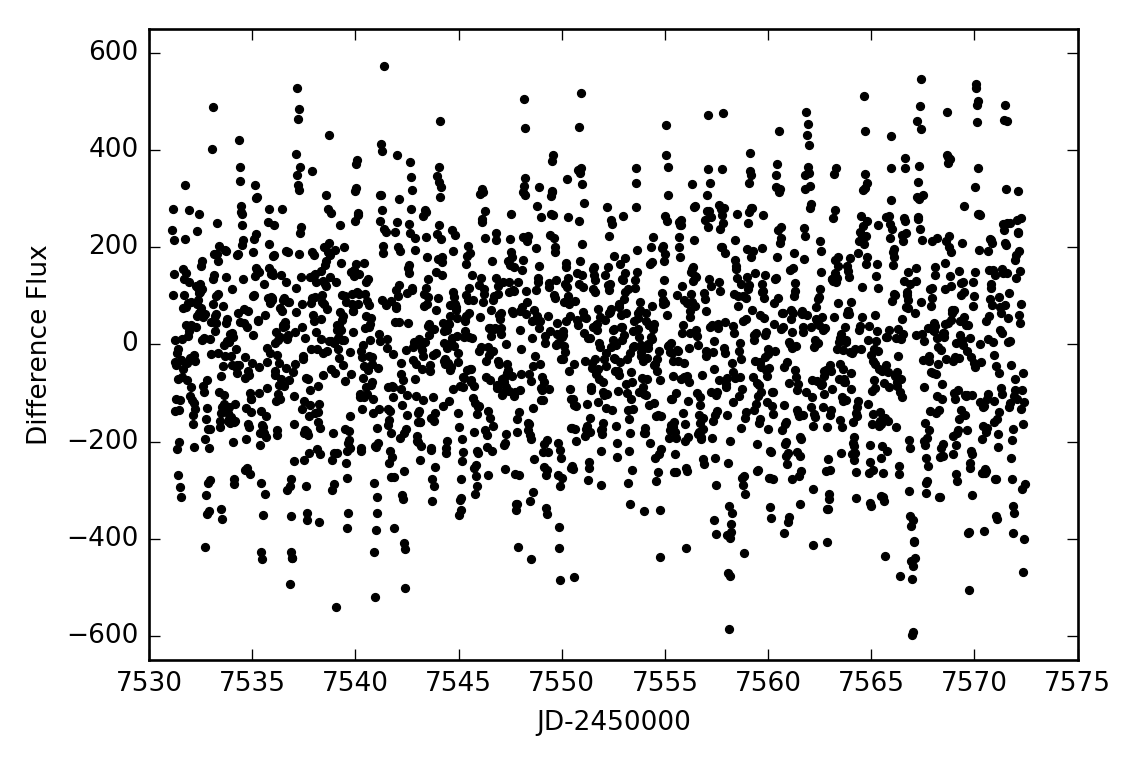}
\includegraphics[width=0.48\textwidth]{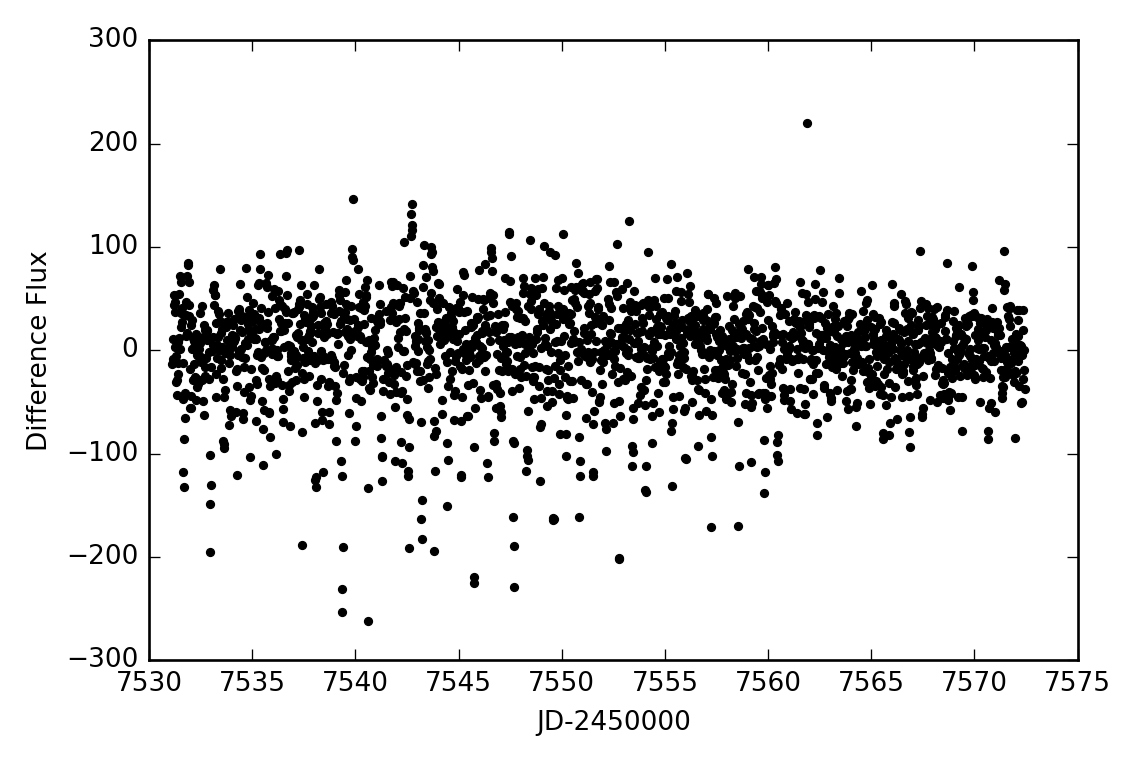}
\includegraphics[width=0.48\textwidth]{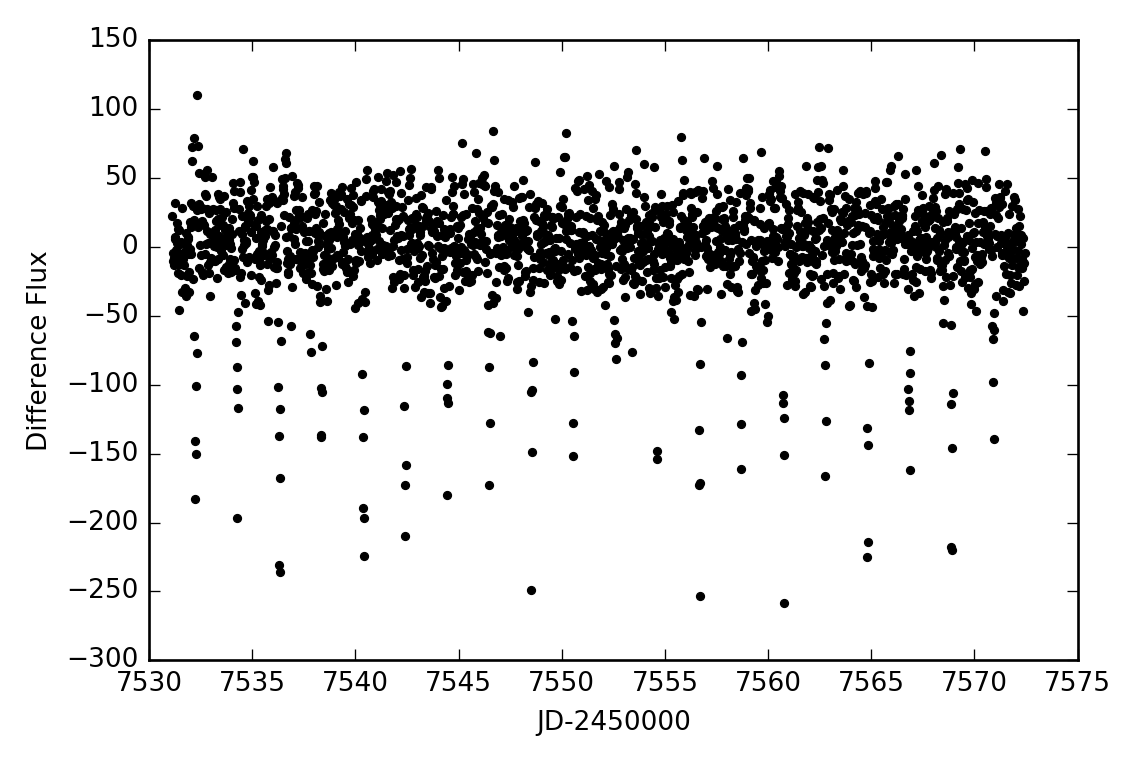}
\includegraphics[width=0.48\textwidth]{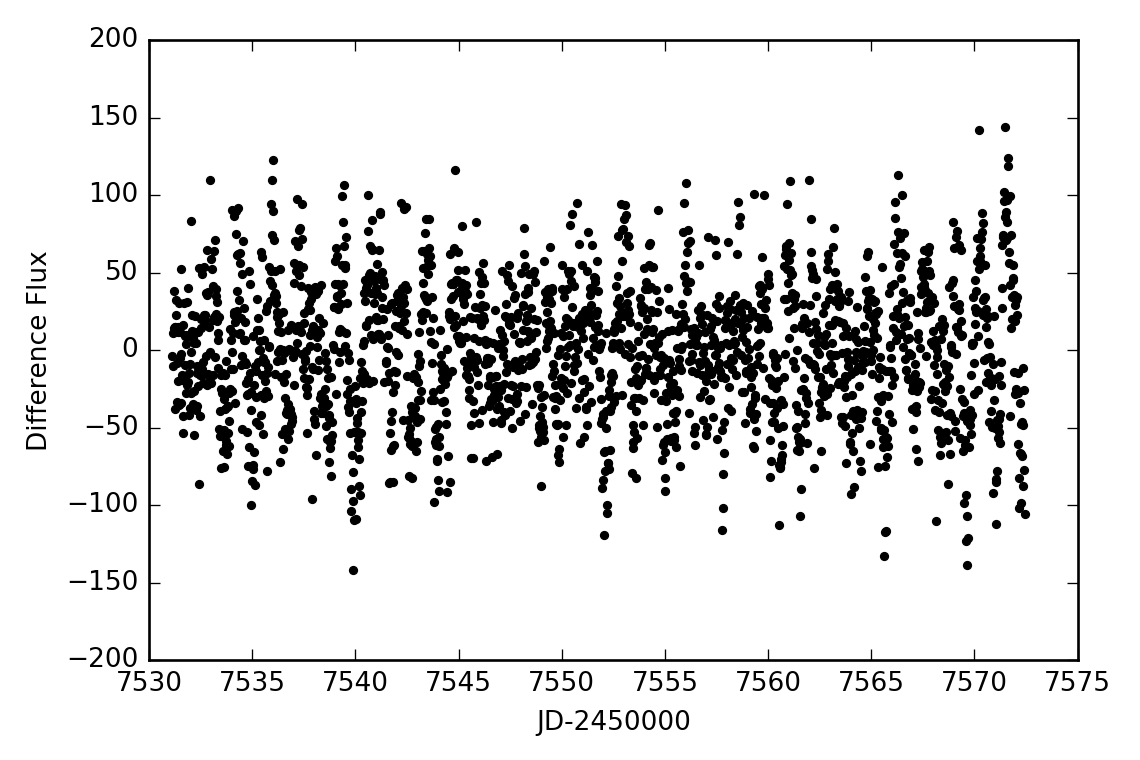}
\includegraphics[width=0.48\textwidth]{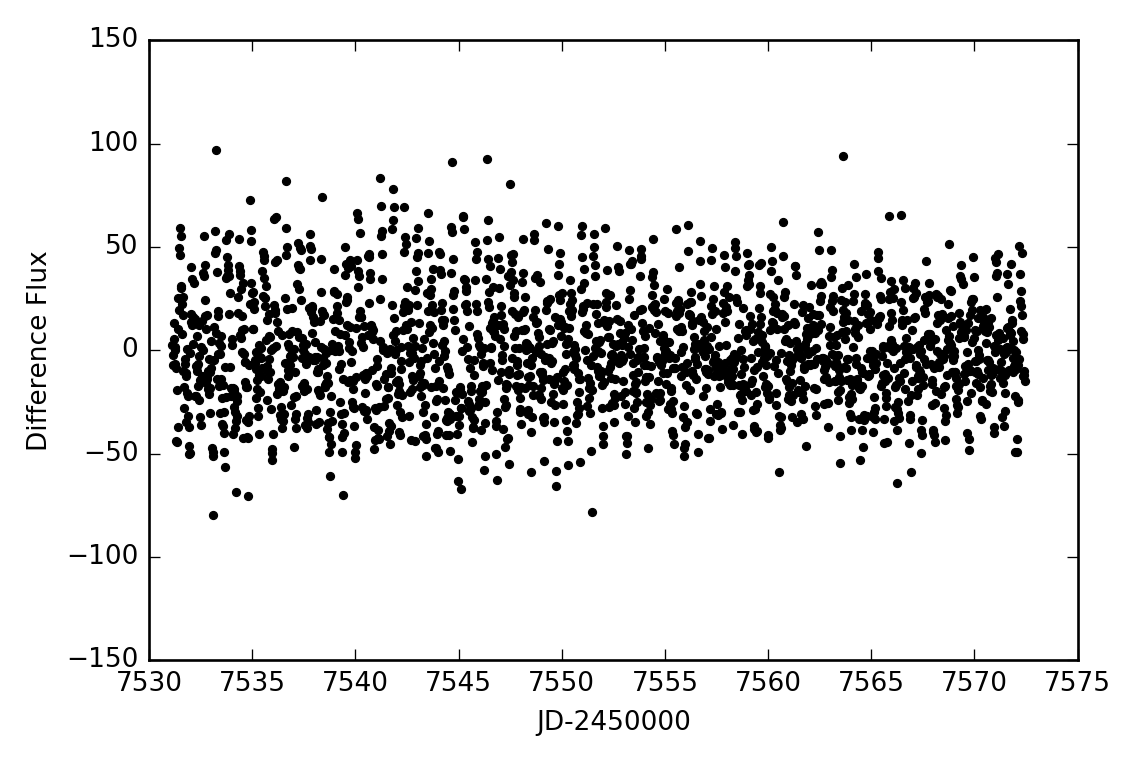}
\includegraphics[width=0.48\textwidth]{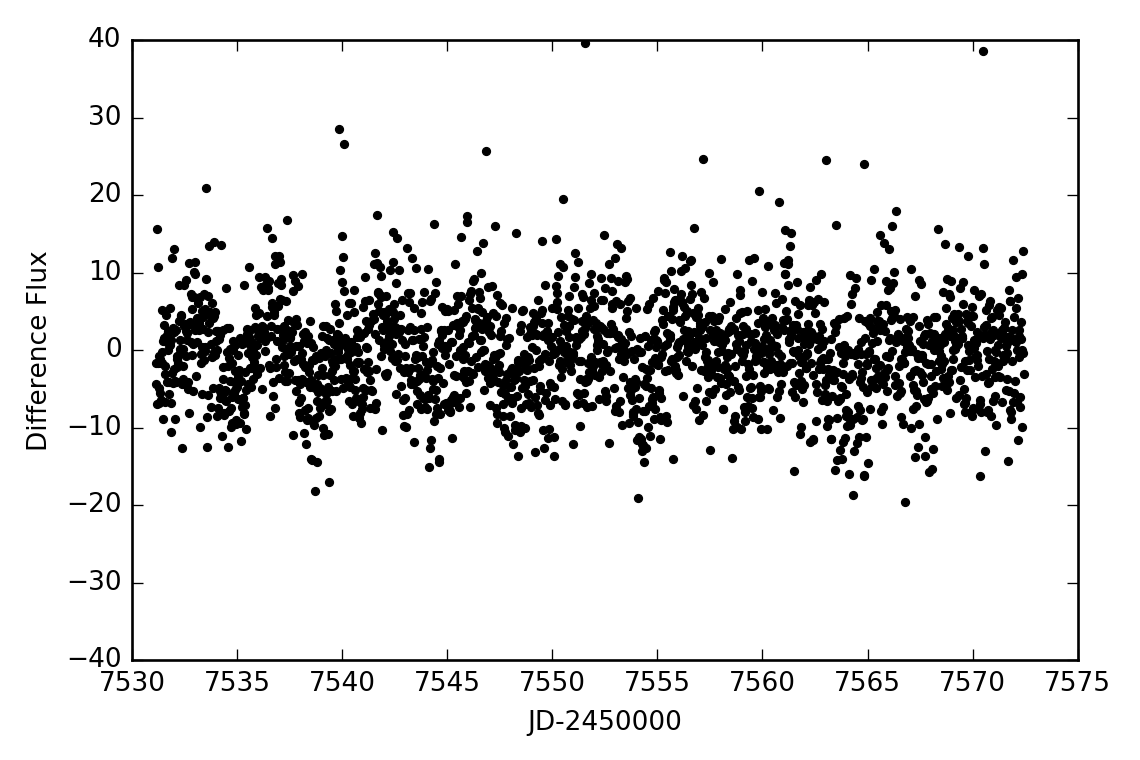}
\end{center}
\caption{
  \label{lightcurve}
  \KTCN\ Data---six light curves extracted from \KTCN\ \epic\ 200069960 by \cpmdiff. 
  Each light curve was generated by co-adding all the difference flux within a $3\times 3$ aperature.
}
\end{figure}

Fig.~\ref{k2c9} shows three snapshots of the data and \cpmdiff\ of different times.
Constant sources in the dense field were almost all cancelled by the \cpm\ prediction, while variables (located at the white crosshairs) were preserved and can even be picked by eye from the difference image.
Variable sources were detected by computing the mean of absolute normalized deviations of the difference images. 
Light curves of six variable sources with high signal-to-noise ratio are presented in Fig.~\ref{lightcurve} as examples. 
Each light curve was constructed with simple aperture photometry, by co-adding all the difference flux within a $3 \times 3$ aperture. 
Note that this is not optimal photometry; it is simply an illustration of what is possible with this method.

\subsection{Limitation of \cpmdiff}\label{limits}
In the above,  \cpmdiff\ was demonstrated on mock and real data. 
It is able to model pointing motion, rotation and PSF variations, such that variable-source detection and photometry can be achieved.
In this Section, we want to push these variations to the limits of what \cpmdiff\ can handle to define the scope of applicability of the method. 
Three experiments are conducted with large variations in pointing, roll and PSF.

\begin{figure}[p]
\begin{center}
\includegraphics[width=0.95\textwidth]{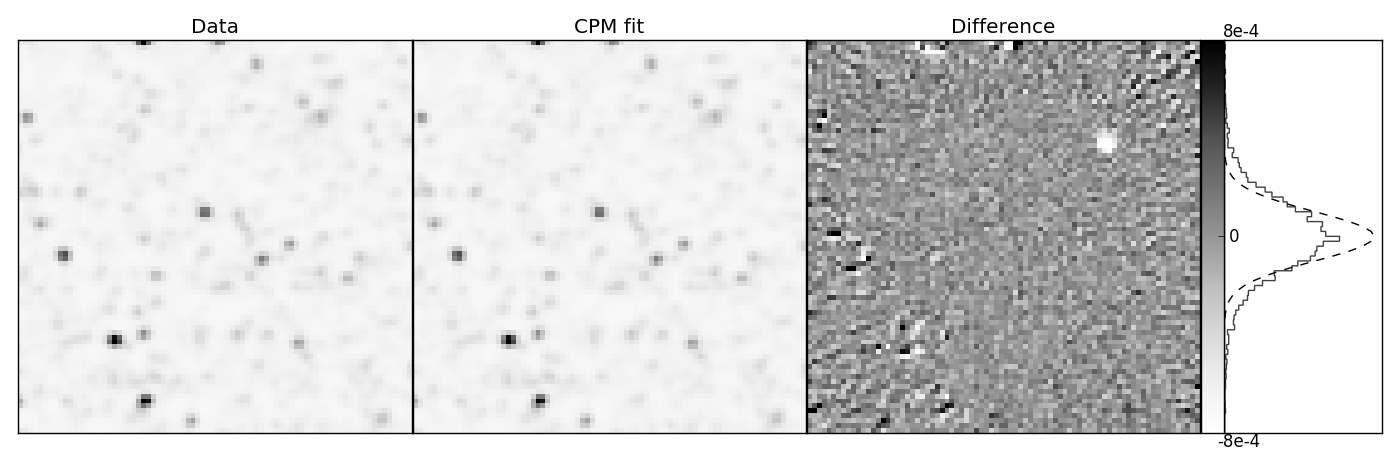}
\includegraphics[width=0.95\textwidth]{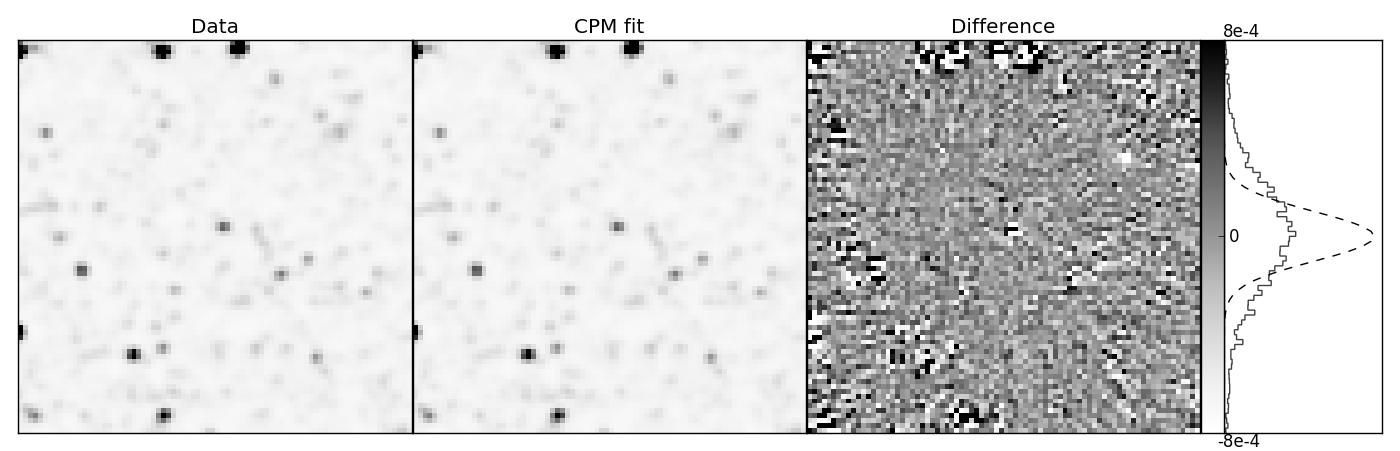}
\includegraphics[width=0.95\textwidth]{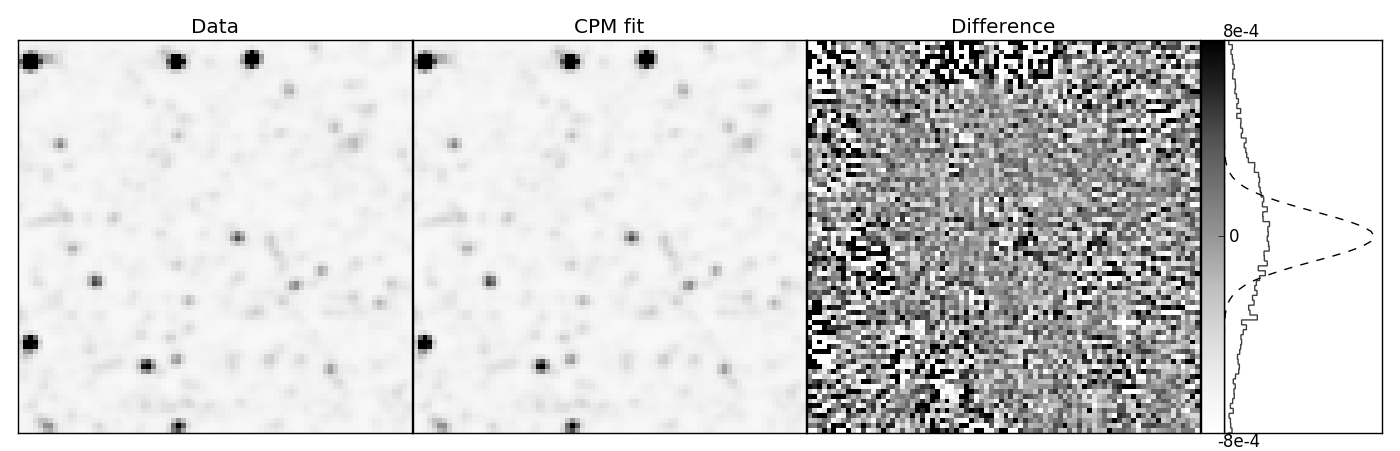}
\end{center}
\caption{
  \label{large_motion}
  Three $80\times 80$ pixel mock data images with the same rotation and PSF variation, but different amplitudes of pointing motion. The amplitude of the pointing motion increases from top to the bottom, which is 2 pixels, 5 pixels and 8 pixels respectively. 
  \emph{Left:} mock data image;
  \emph{Middle:} the prediction of the \cpmdiff;
  \emph{Right:} the relative difference between the data and the prediction, the color bar shows the relative difference; the histogram shows the distribution of the difference and the dashed curve is the photon noise: Gaussian with $\sigma = 10^{-4}$. 
  With the amplitude of the pointing motion increasing, the prediction of the \cpmdiff\ degrades.
}
\end{figure}

\begin{figure}[p]
\begin{center}
\includegraphics[width=0.95\textwidth]{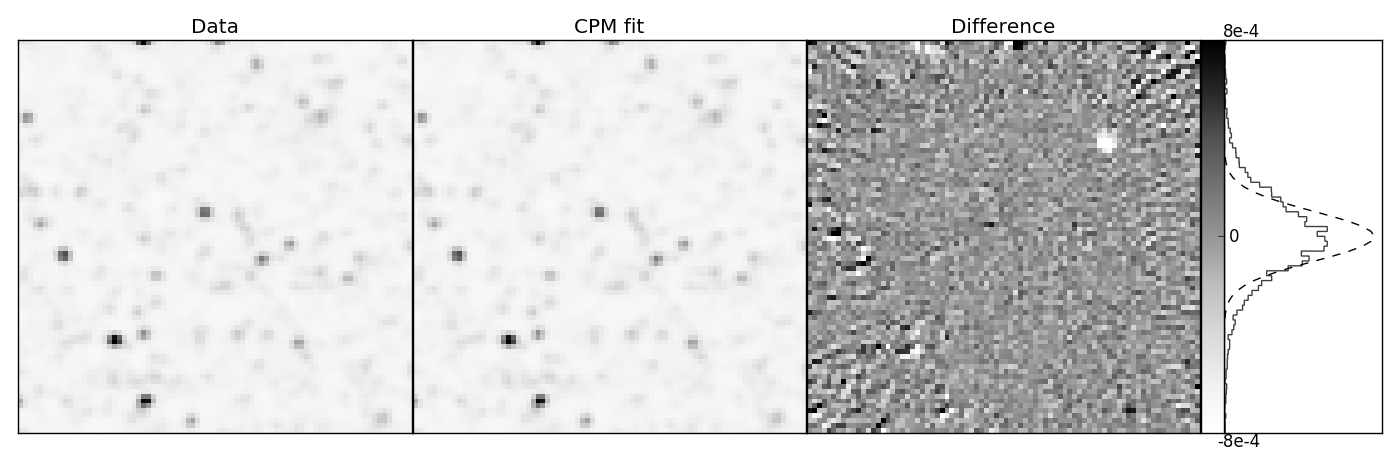}
\includegraphics[width=0.95\textwidth]{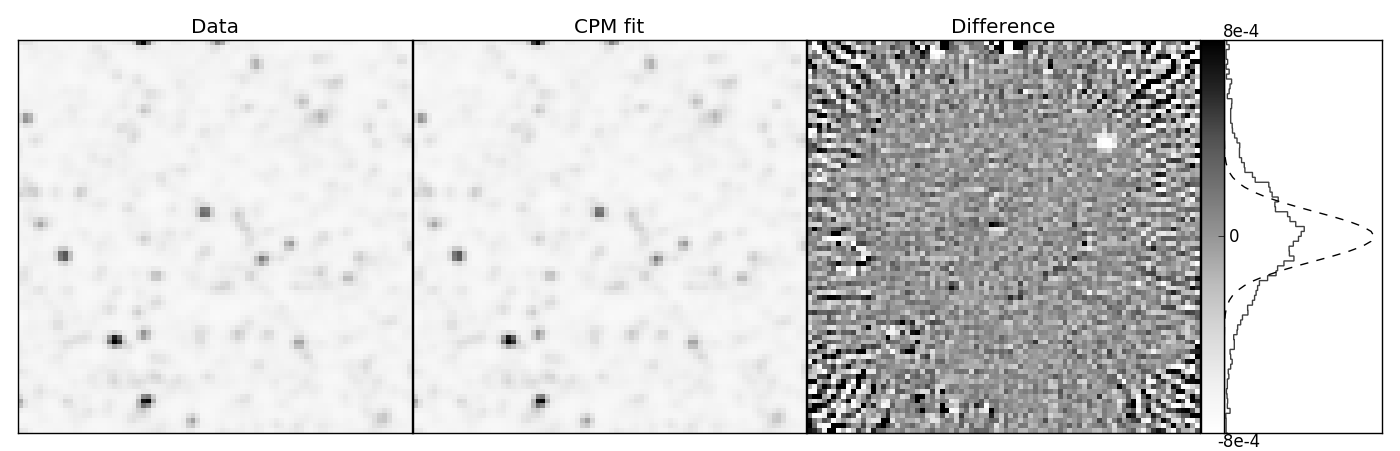}
\includegraphics[width=0.95\textwidth]{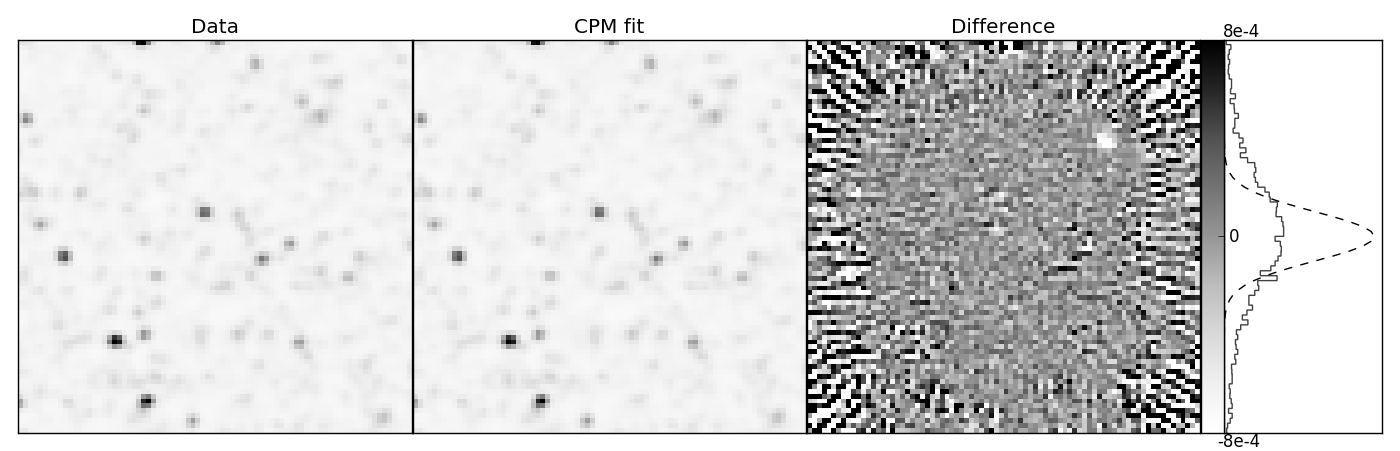}
\end{center}
\caption{
  \label{large_rotation}
  Three $80\times 80$ pixel mock data images with the same pointing motion and PRF variation, but different amplitudes of rotation. The amplitude of rotation increases from top to the bottom, which is 1 deg, 5 deg and 10 deg respectively.
  \emph{Left:} mock data image;
  \emph{Middle:} the prediction of the \cpmdiff;
  \emph{Right:} the relative difference between the data and the prediction, the color bar shows the relative difference; the histogram shows the distribution of the difference and the dashed curve is the photon noise: Gaussian with $\sigma = 10^{-4}$. 
  With the amplitude of rotation increasing, the quality of the difference image drops dramatically in the corner, while almostly remains unchanged in the middle.
}
\end{figure}

\begin{figure}[p]
\begin{center}
\includegraphics[width=0.95\textwidth]{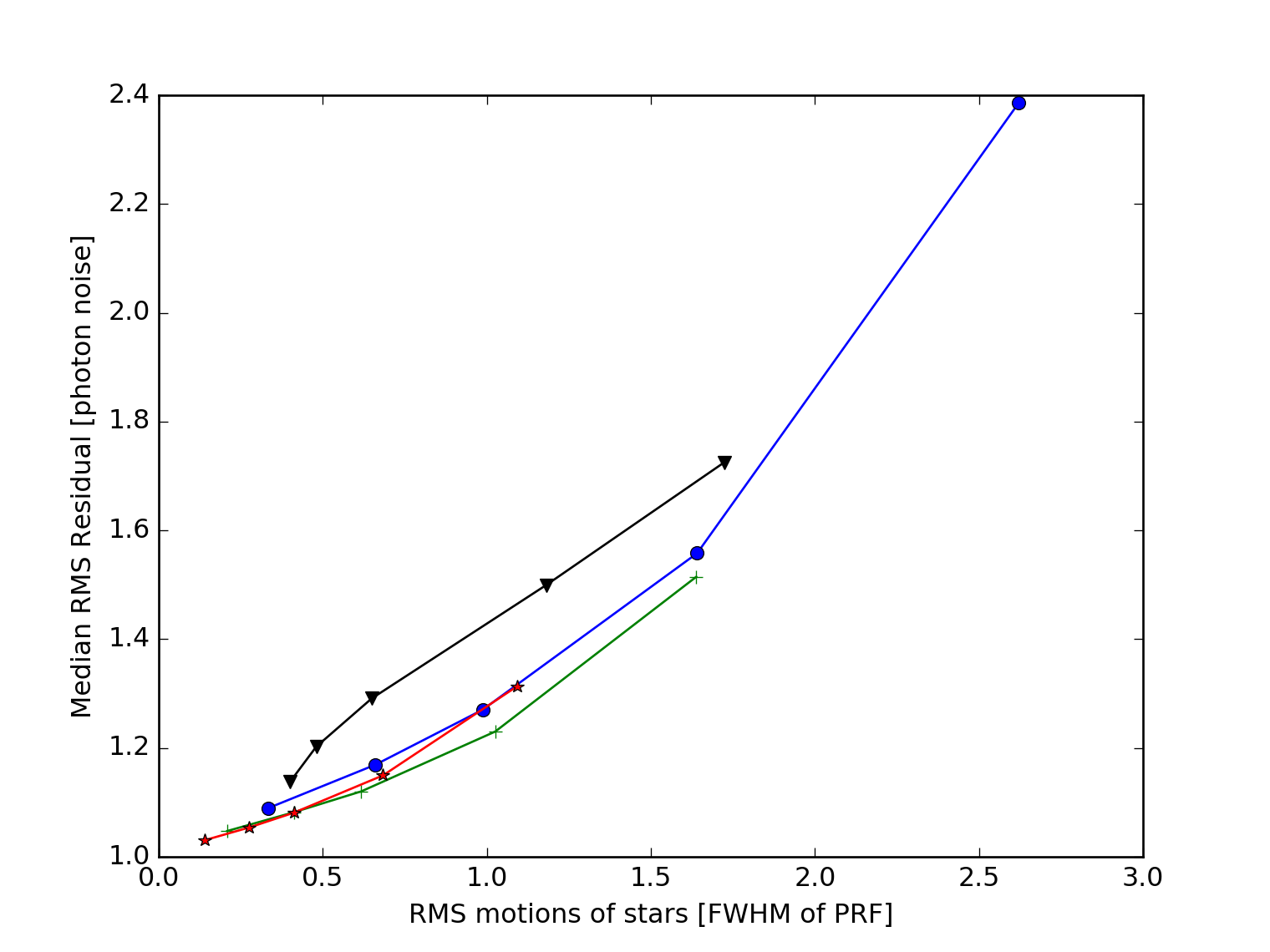}
\end{center}
\caption{
\label{motion_rms}
 Median root-mean-square residual as a function of the root-mean-square motions of stars in unit of FWHM of PRF.
 Different markers and colors indicate different scenarios.
 \emph{Blue dots}: pointing variation with FWHM of PRF = 2.5 pixels;
 \emph{Green cross}: pointing variation with FWHM of PRF = 4 pixels;
 \emph{Red stars}: pointing variation with FWHM of PRF = 6 pixels;
 \emph{Black triangle}: rotation with FWHM of PRF = 2.5 pixels;
}
\end{figure}

First, mock data with the same rotation and PRF variations, but different amplitudes of pointing motion, were tested.
Here both $t_x$ and $t_y$ defined in equation \ref{transformation} were drawn from a uniform distribution ${\mathcal {U}}(0,t_A)$, in which the upper limit $t_A$ defines the amplitude of the pointing motion.
Fig.~\ref{large_motion} shows that with the amplitude of the pointing motion increasing,  the quality of the difference image degrades dramatically, especially around relatively bright sources, because of the additional variabilities introduced by the motion of bright stars.
Similarly, in Fig.~\ref{large_rotation}, mock data with the same pointing and PRF variations, but different amplitudes of rotation were tested by the model.
In each data set, $\theta$ defined in equation \ref{transformation} was drawn from a uniform distribution ${\mathcal {U}}(0,\theta_A)$, in which the upper limit $\theta_A$ defines the maximum amplitude of the rotation.
The corner of the difference image is modelled much worse, since the images are less well aligned in the corner due to the rotation. 
In order to quantitatively study the limitation of the pointing and rotation variation, the quality of each difference image is evaluated by the root-median-square residual (RMS residual) and the overall performance of the model is determined by the median of the RMS Residuals of all the difference images.
The amplitude of the pointing and rotation variations are both translated to the overall motions of stars in unit of FWHM of PRF.
Fig.~\ref{motion_rms} shows the median RMS residual as a function of the RMS motions of stars in different scenarios. 
These four lines in the plot show that image quality degrades with moving stars.
This test also confirms the assumption that the registration should be good to about or better than one PSF width.

\begin{figure}[p]
\begin{center}
\includegraphics[width=0.95\textwidth]{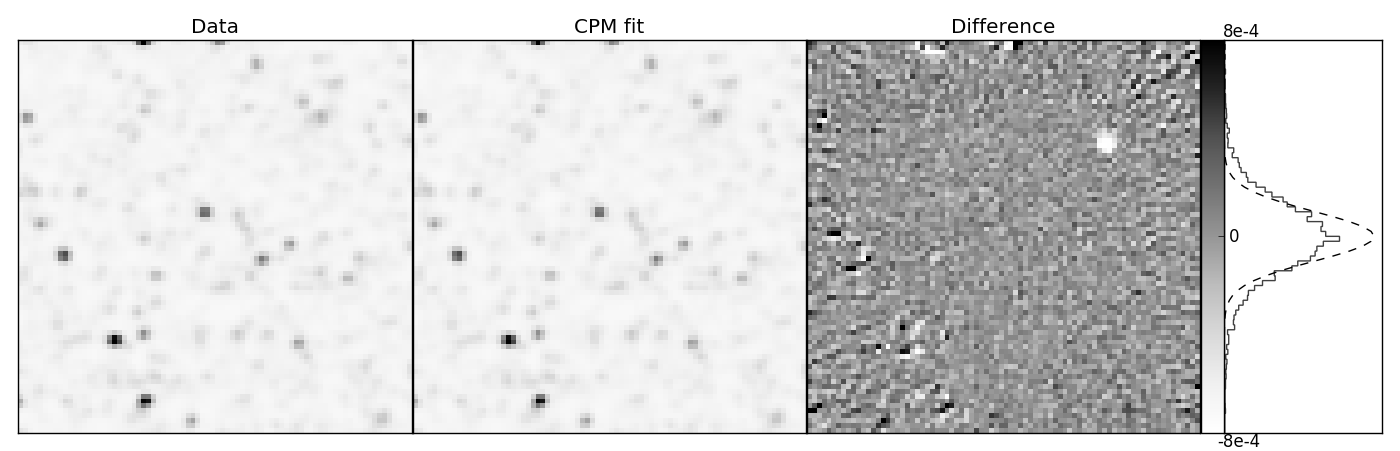}
\includegraphics[width=0.95\textwidth]{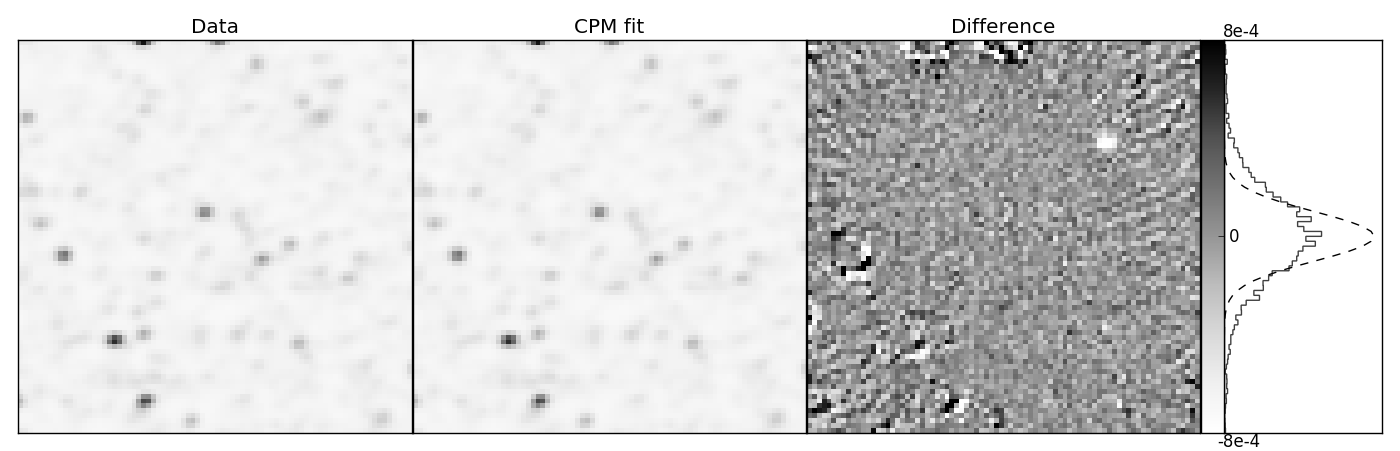}
\includegraphics[width=0.95\textwidth]{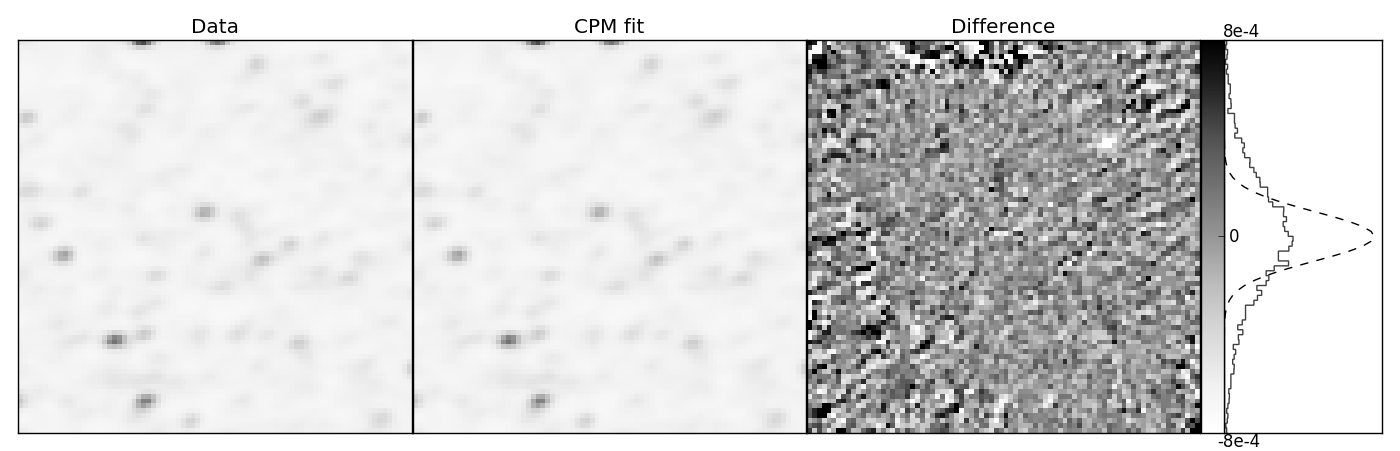}
\end{center}
\caption{
  \label{large_prf}
  Three $80\times 80$ pixel mock data images with the same pointing motion and rotation, but different amplitudes of PRF variation. The amplitude of PRF variation increases from top to the bottom.
  \emph{Left:} mock data image;
  \emph{Middle:} the prediction of the \cpmdiff;
  \emph{Right:} the relative difference between the data and the prediction, the color bar shows the relative difference; the histogram shows the distribution of the difference and the dashed curve is the photon noise: Gaussian with $\sigma = 10^{-4}$. 
  With the amplitude of PRF variation increasing, the quality of the difference image degrades.
}
\end{figure}

\begin{figure}[p]
\begin{center}
\includegraphics[width=0.95\textwidth]{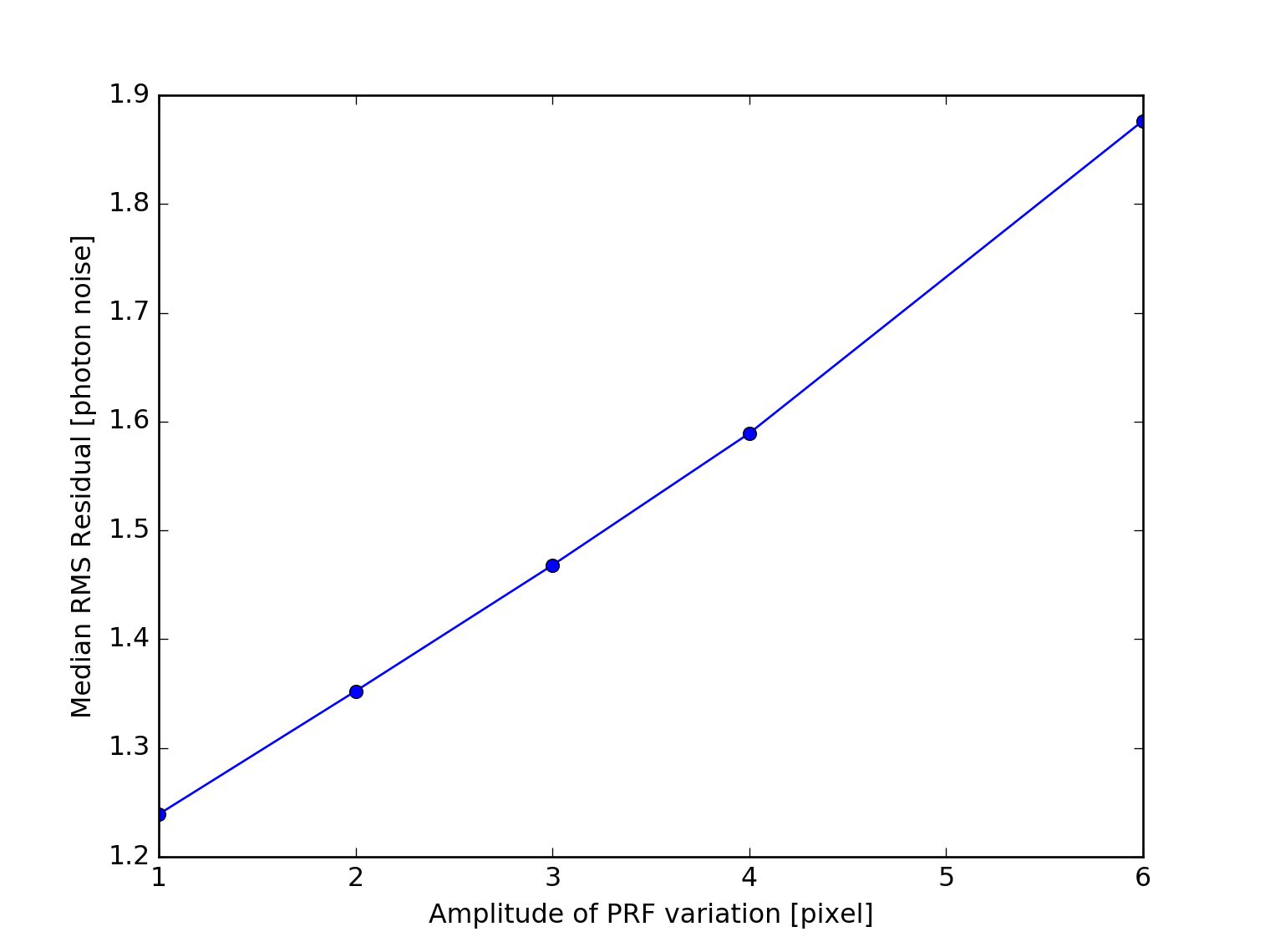}
\end{center}
\caption{
\label{prf_rms}
 Median root-mean-square residual as a function of the amplitude of PRF variation.
}
\end{figure}

Finally, mock data with different amplitudes of PRF variation were tested.
Parameters $f_x$ and $f_y$ defined in equation \ref{prf} were drawn from uniform distribution ${\mathcal {U}}(2,f_{max})$, in which the width of the uniform distribution $f_{max}-2$ defines the amplitude of the PRF variation.
Fig.~\ref{large_prf} and Fig.~\ref{prf_rms} indicate that large PRF variations do degrade the performance of the model, since dramatic changes of the PRF can also significantly change the correlation between pixels.
However, from the experiment, moderate PRF variations are still acceptable. 

\section{Discussion}
We have demonstrated that \cpmdiff\ is capable of predicting images with precision close to the photon noise, while preserving the variable sources. We have shown this with simulated data comparable to typical space-based and ground-based data with moderate pointing, rotation and PSF variations. 
Here we list few reasons \cpmdiff\ may be preferred over other methods:

\begin{enumerate}
\item All classical difference imaging approaches require precise image registration as a first/prior step. 
Any astrometric alignment error between image frames will lead to imperfection after subtraction.
In comparison, the precision requirement of registration is rather relaxed in \cpmdiff. 
As demonstrated in Section \ref{limits}, with pointing variation smaller than $\sim$1 FWHM of PSF, precision close to photon noise can still be achieved by our method.  

\item All classical difference imaging approaches require direct or indirect information about the PSF. 
Any imperfection of the PSF(or difference of PSFs) estimation will lead to subtraction residuals.
\cpmdiff\ avoids PSF estimation, since no convolution kernel is used.

\item For photometry, a de-trending process may be required after image subtraction in the classical approach to account for extra variabilities induced by intra-and inter-pixel variations.
In comparison, \cpmdiff\ mitigates these effects internally.
\end{enumerate}

Apart from these benefits of the method, we emphasize that \cpmdiff\ has its own scope of applicability.
The method requires enough images and pixels to optimize the model, so it can not be used in a situation where there are few images or the size of the field is too small. 
From Section \ref{limits}, we learned that although \cpmdiff\ does not require perfect knowledge of pointing or any knowledge of the PSF, it is important to ensure the registration to be good within $\sim$1 FWHM of PSF for best performance.
In addition, PSF variations between frames should also be moderate, so that the dependence between pixels are approximately invariant over time.

Although \cpmdiff\ performs well for the examples in this paper,  in order to fully exploit the potential of the method, it is still important to find a good set of hyper-parameters and optimize the selection of predictors as mentioned in Section \ref{method}. 
In the original \cpm, there are two hyper-parameters---the number of predictor pixels N and the strength of the L2-regularization $\lambda_a$, which can be optimized by cross-validation. 
However, the selection or ranking of the predictor pixels have been heuristically defined here by either using the closest pixels in space or in brightness.
This intuitive setting is neither fully tested nor optimal, so exploring how to filter or select predictor pixels is necessary to achieve better performance of the model.
There are many ways worth trying to improve the selection of predictors.
For example, one can filter the predictor pixels by variability of the pixel.
Ideally, only quiet pixels that do not contain much variability from variable stars should be included in the predictor set, since variable predictor pixels may distort the target pixel or introduce artefacts.
Therefore if possible, all the known astrophysically variable pixels should be excluded from the predictor set. 
Another way to improve can be running \cpm\ with L1-regularization and a large set of pixels. 
Since L1-regularization leads to sparsity of the coefficients, it can filter the pixels that do not contribute in the fitting. 
Other feature reduction and extraction methods such as PCA or adding higher order components might also possibly improve the set of the predictors and further enhance the performance of the model. 

To conclude, the proposed new approach of image differencing (\cpmdiff) is capable of variability search in time-domain imaging from both space-based and ground-based data.
The performance of the method can still be further enhanced by optimizing the hyper-parameters or exploring possible ways to improve the predictor set.
We hope \cpmdiff\ can be useful and achieve more scientific results in the future survey such as \project{LSST} and \project{TESS}.

\acknowledgements
It is a pleasure to thank
  Federica B. Bianco (NYU)
  and
  Dustin Lang (Toronto)
for valuable comments and suggestions.
This research was partially supported by
  NASA (grant NNX12AI50G),
  NSF (grants IIS-1124794, AST-1312863, AST-1517237),
  and NG Next.
The data analysis presented in this article was partially performed on computational resources at NYU HPC.
This research made use of the NASA Astrophysics Data System.

\clearpage
\bibliography{cdi}
\clearpage

\end{document}